\documentclass[]{IEEEapm}
\usepackage{graphicx}  
\usepackage{dcolumn}   
\usepackage{bm}        
\usepackage{verbatim}   
\usepackage{graphicx}
\usepackage{epstopdf}
\usepackage{array}
\usepackage{cite}
\usepackage{psfrag}
\usepackage{booktabs}
\usepackage{footnote}
\usepackage[most]{tcolorbox}
\usepackage{epsfig}
\usepackage{subfigure}
\usepackage{psfrag}
\usepackage{amssymb}
\usepackage{amsmath,bm}
\usepackage{lipsum}
\usepackage{mathtools}
\usepackage{cuted}
\usepackage[utf8]{inputenc}
\usepackage{lipsum} 

\newcolumntype{P}[1]{>{\centering\arraybackslash}p{#1}}

\usepackage[framemethod=TikZ]{mdframed}
\usepackage{lipsum}
\jmonth{November}
\pubyear{2024}

\begin{document}
	
	\title{Finite-Difference Time-Domain Simulation of Wave Transmission Through Space-Time-Varying Media}
	
	\author{Sajjad Taravati$^{1}$,~\IEEEmembership{Senior Member,~IEEE}, Ahmed A Kishk$^{2}$,~\IEEEmembership{Life Fellow,~IEEE},  
		and George V Eleftheriades$^{3}$,~\IEEEmembership{Fellow,~IEEE}}
	
	\affil{$^{1}$Faculty of Engineering and Physical Sciences, University of Southampton, Southampton SO17 1BJ, United Kingdom\\
		$^{2}$Department of Electrical and Computer Engineering, Concordia University, Montreal, QC H3G 1M8, Canada\\
		$^{3}$Department of Electrical and Computer Engineering, University of Toronto, Toronto M5S 3H7, Canada\\
		(e-mail: S.Taravati@soton.ac.uk)}

	\maketitle
	
	
	\begin{receivedinfo}%
	\end{receivedinfo}
	
	\begin{abstract}
A comprehensive study on the Finite Difference Time Domain (FDTD) numerical modelling of space- and time-varying media is presented. We investigate the dynamic behavior of oblique incidence of both TM and TE electromagnetic fields on space-time-modulated gratings. In their general form, these media have electrical permittivity, magnetic permeability, and electrical conductivity modulated across both space and time. We provide the FDTD schemes, the Courant-Friedrichs-Lewy (CFL) stability condition and equations for both TM and TE wave illuminations of space-time-varying slabs, along with the boundary conditions for the FDTD scheme. Furthermore, with engaging illustrative examples, we showcase the versatility and applications of space-time-varying media.
	\end{abstract}
	

\section{Introduction}
\label{sec:introduction}

\IEEEPARstart{T}{ime-varying} metamaterials and metasurfaces have spurred significant attention due to their wide-ranging applications in modern wireless communication systems, photonics, wave engineering, radar technologies, and beyond~\cite{Taravati_Kishk_MicMag_2019,wang2022imaging,Taravati_PRAp_2018,saikia2022time,Taravati_Kishk_PRB_2018,taravati_PRApp_2019,Taravati_Kishk_TAP_2019,Taravati_AMA_PRApp_2020,sabri2021broadband,taravati2020full}. These structures exhibit dynamic properties characterized by the modulation of electrical permittivity, magnetic permeability, and electrical conductivity across both space and time~\cite{wang2021space,taravati2021programmable,wan2021nonreciprocal,Taravati_PRB_SB_2017,taravati20234d,amra2024linear,Taravati_NC_2021,taravati2024reflective,valizadeh2024analytical}. Understanding their behavior is crucial for designing advanced devices and systems with enhanced functionality and performance. Analyzing wave propagation in these media and the associated physical phenomena necessitates a deep comprehension of electrodynamics, including Lorentz transformations~\cite{Lorentz1937electromagnetic}, electromagnetic wave propagation, and wave-vector diagrams in space-time periodic media~\cite{elachi1972electromagnetic,elachi1976waves,taravati_PRB_2017,taravati20234d,darvish2024modern,oudich2023tailoring,huidobro2019fresnel,stewart2017finite}, as well as unique analytical implications~\cite{solis2021functional,pacheco2021temporal}. Space-time metasurfaces can be realized at microwave~\cite{saikia2019frequency,cardin2020surface,Taravati_ACSP_2022,kumar2024multi} and optical frequencies~\cite{salary2020time,sabri2021broadband,sisler2024electrically} for various functionalities, including nonreciprocal transmission~\cite{taravati2020full,cardin2020surface,kord2020microwave,Taravati_ACSP_2022,taravati2023nonreciprocal}, target recognition~\cite{wang2023pseudorandom}, isolators~\cite{Taravati_PRB_SB_2017,Taravati_AMTech_2021}, temporal aiming~\cite{pacheco2020temporal}, pure frequency conversion~\cite{taravati2021pure,Taravati_PRB_Mixer_2018}, static-to-dynamic field conversion~\cite{mencagli2022static}, circulators~\cite{Alu_NPH_2014,dinc2017millimeter,taravati2022low}, parametric amplification~\cite{zhu2020tunable,pendry2020new}, multiple access secure communication systems~\cite{taravati_PRApp_2019,sedeh2021active}, nonreciprocal antennas~\cite{zang2019nonreciprocal,zang2019nonreciprocal2}, coding metasurfaces~\cite{zhang2019dynamically}, and multifunctional operations~\cite{Taravati_LWA_2017,wang2019multifunctional,Taravati_AMA_PRApp_2020}.

This Tutorial presents a Finite Difference Time Domain (FDTD) numerical simulation scheme for modeling space- and time-varying media. We apply the FDTD method to simulate electromagnetic wave scattering from space-time modulated media. These media exhibit properties that vary both spatially and temporally, introducing additional complexity into the simulation. The time-varying permittivity $\epsilon(z,t)$, permeability $\mu(z,t)$, and conductivity $\sigma(z,t)$ of the medium must be incorporated into the updating equations to model the electromagnetic behavior accurately. Detailed equations and methodologies for both TE and TM wave illuminations of space-time-varying slabs are provided to facilitate fast and accurate simulation of these complex structures. By offering a versatile and efficient tool through our proposed FDTD scheme, researchers and engineers can gain access to a platform for investigating the dynamic behavior of space-time-varying electromagnetic components. This advancement enables the rapid development and optimization of novel electromagnetic devices and antennas with tailored functionalities, fostering progress in communication systems, radar technologies, and various engineering applications.

The paper is organized as follows. Section~\ref{sec:categ} provides an overview of different categories of space, time, and space-time metamaterials and metasurfaces, detailing their unique properties. Section~\ref{sec:theo} introduces the fundamentals of FDTD simulation for space-time-varying media, including the Courant stability factor, and the behavior of the refractive index and electromagnetic fields in these media. Section~\ref{sec:TM} details the FDTD simulation scheme and electromagnetic field update equations for oblique TM wave incidence and transmission through space-time-varying media, accompanied by illustrative examples and results. Section~\ref{sec:TE} describes the FDTD simulation scheme and electromagnetic fields update equations for oblique TE wave incidence and transmission through space-time-varying media, also followed by illustrative examples. Finally, Sec.~\ref{sec:conc} concludes the paper, summarizing key findings and implications.

\section{Time-Varying Metamaterials and Metasurfaces}\label{sec:categ}

\begin{figure}
	\begin{center}
		\includegraphics[width=0.5\columnwidth]{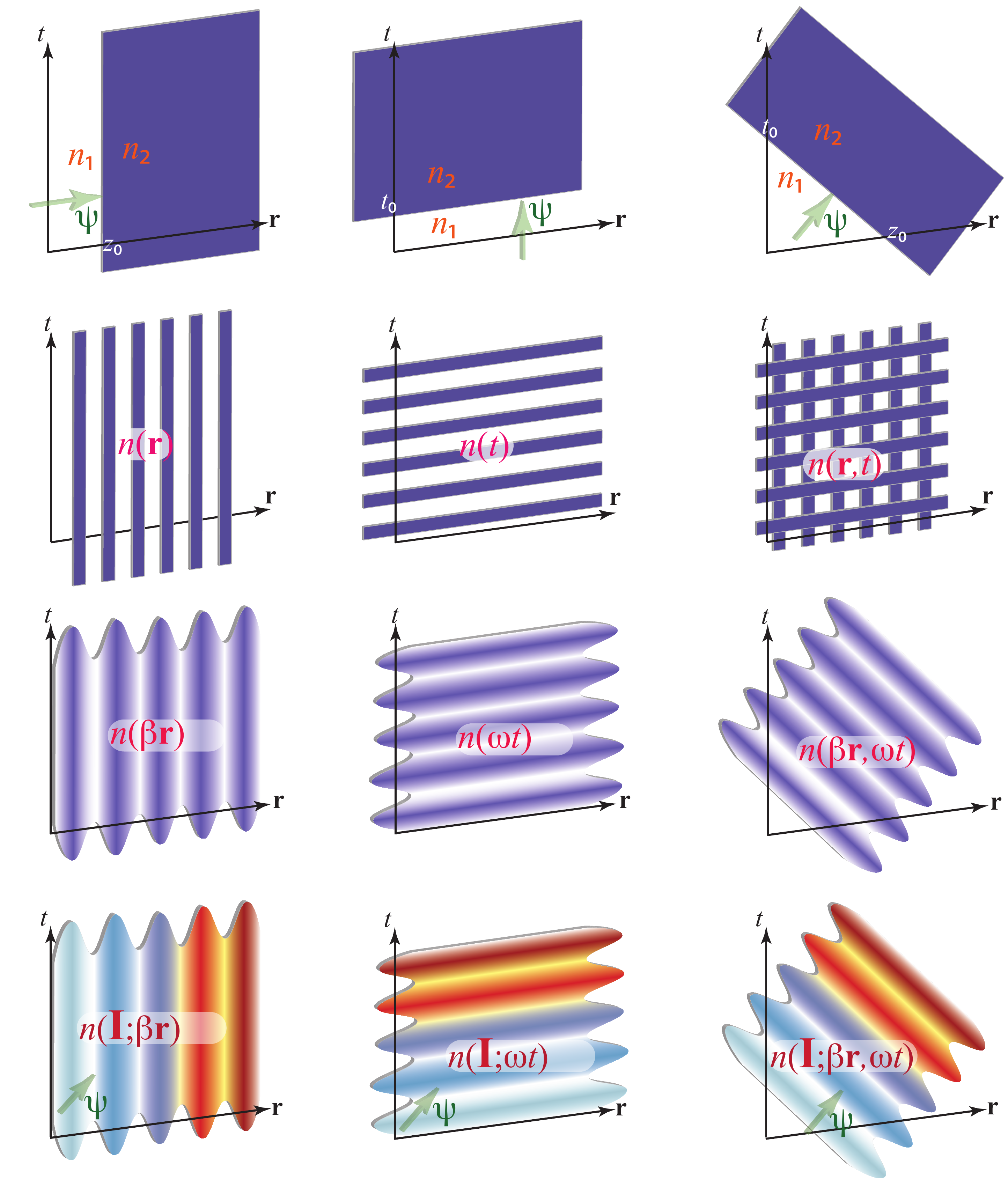}
		\caption{Representation of Inhomogeneous Metamaterials and Metasurfaces in Four-Dimensional Minkowski Space: 1st column: Spatially inhomogeneous (variations in space). 2nd column: Temporally inhomogeneous (variations in time). 3rd column: Spatiotemporally inhomogeneous (variations in both space and time).}
		\label{Fig:categ}
	\end{center}
\end{figure}

Figure~\ref{Fig:categ} illustrates the Minkowski space of 12 distinct categories of metamaterials and metasurfaces (MMs $\&$ MSs), classified based on the dependencies of their refractive index $n$ on space and time. This includes, spatial abrupt discontinuity $n_{\textbf{r}>\textbf{r}_0}=n_2$, temporal abrupt discontinuity $n_{t>t_0}=n_2$, spatiotemporal abrupt discontinuity $n_{\{\textbf{r}>\textbf{r}_0;t>t_0\}}=n_2$, spatial photonic crystals $n(\textbf{r})$, temporal photonic crystals $n(t)$, spatiotemporal photonic crystals $n(\textbf{r},t)$, linear spatially-modulated media $n(\beta \textbf{r})$, linear temporally-modulated media $n(\omega t)$, linear spatiotemporally-modulated media $n(\beta \textbf{r},\omega t)$, nonlinear spatially-modulated media $n(\textbf{I};\textbf{r})$, nonlinear temporally-modulated media $n(\textbf{I};t)$, and nonlinear spatiotemporally-modulated media $n(\textbf{I};\textbf{r},t)$.

Minkowski space, also known as spacetime, is a cornerstone of modern physics, particularly in Einstein's special relativity. This mathematical framework unifies space and time into a single, four-dimensional fabric, where both are treated equally. Unlike the classical view of separate three-dimensional space and one-dimensional time, Minkowski space integrates them into a cohesive whole. A key concept in this framework is the idea of events, which are points in spacetime defined by specific spatial coordinates ($\textbf{r}$) and time coordinates ($t$). These coordinates ($\textbf{r}$ and $t$) are crucial for pinpointing the location and timing of events, enabling a precise and thorough understanding of physical phenomena. In the realm of MMs $\&$ MSs, Minkowski space becomes particularly relevant when considering the behavior of electromagnetic waves. By visualizing these variations within the framework of Minkowski space, researchers can gain deeper insights into how these materials interact with electromagnetic waves, paving the way for innovative applications in fields such as wireless communications, biomedicine and quantum technologies.
 
Linear space-modulated MMs $\&$ MSs represented by $n(\textbf{r})$ are materials whose refractive index varies spatially but does not depend on the intensity of the optical signal. These materials exhibit spatial modulation of the refractive index, which can be controlled to manipulate the propagation of light. Nonuniform transmission lines serve as examples of linear space-modulated media. Over the past decade, linear materials with dependence on time $n(t)$ and space-time $n(\textbf{r},t)$ have been widely studied. Taking advantage of phase-engineered $n(t)$ MMs $\&$ MSs, several enhanced efficiency apparatuses and novel functionalities have been reported, including nonreciprocal-beam MS, ideal frequency converter MS, low-noise isolators and circulators, and enhanced resolution imaging. The $n(r,t)$ MMs $\&$ MSs break time reversal symmetry by employing unidirectional space-time modulation. The research in this area has led to innovative designs for non-reciprocal and multifunctional components, including one-way beam-splitter~\cite{Taravati_Kishk_PRB_2018}, antenna-mixer-amplifier~\cite{Taravati_AMA_PRApp_2020}, mixer-duplexer-leaky-wave-antenna~\cite{Taravati_LWA_2017}, pure frequency converter~\cite{Taravati_PRB_Mixer_2018}, and magnet-less isolators~\cite{taravati_PRB_2017,Taravati_PRB_SB_2017}. This breakthrough enables the integration of signal processing and antenna functionality within a singular MM framework, heralding a new class of multifunctional systems. 

Nonlinear materials $n(\textbf{I})$ like superconductors show properties that change with current intensity. The Josephson effect, observed in Josephson junctions made from materials like YBCO or NbTi, relies on the quantum tunneling of Cooper pairs influenced by this current. This enables zero voltage current flow across the junction, which is pivotal for highlighting the essential role of precise current management in SC applications. SCs, key since 1911, facilitate major technological advances across multiple fields. Nonlinear space-modulated MMs $\&$ MSs represented by $n(\textbf{I};\textbf{r})$ are materials whose refractive index varies both spatially and nonlinearly with the intensity of the optical signal. These materials combine spatial modulation with nonlinear optical effects, enabling complex control over light propagation and interactions. Nonlinear Bragg gratings and photonic crystals with engineered nonlinearities are examples of nonlinear space-modulated MMs $\&$ MSs. These gratings consist of periodic variations in the refractive index along the length of an optical fiber, and the intensity-dependent refractive index allows for nonlinear phenomena such as soliton formation and optical switching. 

Nonlinear space-time-modulated media are represented by $n(\textbf{I};t)=n_0(t)+n_2(t)\textbf{I}(t)$ and $n(\textbf{I};\textbf{r},t)=n_0(\textbf{r},t)+n_2(\textbf{r},t)\textbf{\textbf{I}}(\textbf{r},t)$. Here, $n_0(\textbf{r},t)$ represents the spatial and temporal variation of the linear refractive index, $n_2(r,t)$ represents the spatial and temporal variation of the nonlinear refractive index coefficient, and $\textbf{I}(\textbf{r},t)$ represents the spatial and temporal variation of the intensity of the signal. The $n(\textbf{I};\textbf{r},t)$ metamaterials (MMs) and metasurfaces (MSs) represent the most advanced and comprehensive class of space-time-varying media. For example, space-time-varying Josephson metasurfaces have recently been proposed for wave engineering at the quantum level, demonstrating compatibility with millikelvin-temperature quantum technologies~\cite{taravati2024spatiotemporal}.

\section{Theory}\label{sec:theo}

The FDTD method is a robust and versatile numerical technique widely used to solve Maxwell's equations in the time domain. Introduced by K. S. Yee in 1966, the FDTD method discretizes both spatial and temporal domains using a grid, transforming continuous differential equations into a set of algebraic equations that can be solved iteratively. This method is particularly powerful for simulating complex electromagnetic phenomena in materials with varying properties over space and time. In the FDTD method, the spatial domain is divided into a grid with cell dimensions $\Delta x$, $\Delta y$, and $\Delta z$, and the time domain is discretized with a time step $\Delta t$. The electric and magnetic fields are updated at each time step using finite-difference approximations of the derivatives. The updating equations for the electric and magnetic fields can be derived using central-difference approximations.

The stability of the FDTD method is governed by the Courant-Friedrichs-Lewy (CFL) condition, which ensures that the numerical solution remains stable. The general form of the CFL condition for a FDTD scheme in a homogeneous medium is given by $v \Delta t/\Delta \textbf{r} \leq 1$, where $v$ is the speed of light in the medium. For a medium with a space-time varying refractive index, the stability criterion can be influenced by the variations in the refractive index. Therefore, the stability condition must take into account the maximum value of the refractive index over space and time to ensure stability throughout the simulation domain. The CFL condition reads $\max\left(|v(\textbf{I};\textbf{r},t)|\right) \Delta t /  \Delta \textbf{r}  \leq 1$, where $\max\left(|v(\textbf{I};\textbf{r},t) |\right)$ represents the highest speed of light in the medium, and the CFL condition ensures that the time step $\Delta t$ is small enough to accommodate this slowest propagation speed throughout the entire simulation domain. Therefore, the CFL condition reads
\begin{equation}
	\Delta t \leq \frac{1}{\max\left(|v(\textbf{I};\textbf{r},t)| \right) \sqrt{\frac{1}{\Delta x^2} + \frac{1}{\Delta y^2} + \frac{1}{\Delta z^2}} \cdot },
\end{equation}
This condition implies that the time step $\Delta t$ must be sufficiently small relative to the spatial discretization to ensure stability. The constitutive relations in the medium are
	\begin{subequations}
\begin{equation}
	\mathbf{D} = \epsilon_\text{m}(\textbf{I};\textbf{r},t) \mathbf{E}
\end{equation}
\begin{equation}
	\mathbf{B} = \mu_\text{m}(\textbf{I};\textbf{r},t) \mathbf{H}
\end{equation}
\begin{equation}
	\mathbf{J} = \sigma_\text{m}(\textbf{I};\textbf{r},t) \mathbf{E}
\end{equation}
	\end{subequations}
		
Thus, the refractive index \(n(\textbf{I};\textbf{r},t)\) of the medium is given by:
\begin{equation}
	n(\textbf{I};\textbf{r},t) = c \sqrt{ \mu_\text{m}(\textbf{I};\textbf{r},t) \left( \epsilon_\text{m}(\textbf{I};\textbf{r},t) - i \frac{\sigma_\text{m}(\textbf{I};\textbf{r},t)}{\omega} \right) }
\end{equation}

The Faraday’s law of induction gives the fundamental Maxwell’s equations in differential form for a medium as
	\begin{subequations}
		Faraday's law of induction:
		\begin{equation}\label{eqa:Max1}
			\nabla\times\mathbf{E}_\text{m} =-\dfrac{\partial [\mu_\text{m}(\textbf{I};\textbf{r},t) \mathbf{H}_\text{m} ]}{\partial t}
		\end{equation}
		and Ampère's law with Maxwell's addition:
		\begin{equation}\label{eqa:Max2}
			\nabla\times\mathbf{H}_\text{m}=\dfrac{\partial[\epsilon_\text{m}(\textbf{I};\textbf{r},t) \mathbf{E}_\text{m} ]}{\partial t}+\sigma_\text{m}(\textbf{I};\textbf{r},t) \mathbf{E}_\text{m},
		\end{equation}
	\end{subequations}
	where $\mathbf{E}_\text{m}$ and $\mathbf{H}_\text{m}$ are the electric and magnetic fields, respectively, while $\epsilon_\text{m}$, $\mu_\text{m}$, and $\sigma_\text{m}$ are the medium's time and space-dependent permittivity, permeability, and conductivity.

\section{FDTD Scheme and Equations for TM Wave Excitation}\label{sec:TM}

\begin{figure}
	\begin{center}
		\includegraphics[width=0.4\columnwidth]{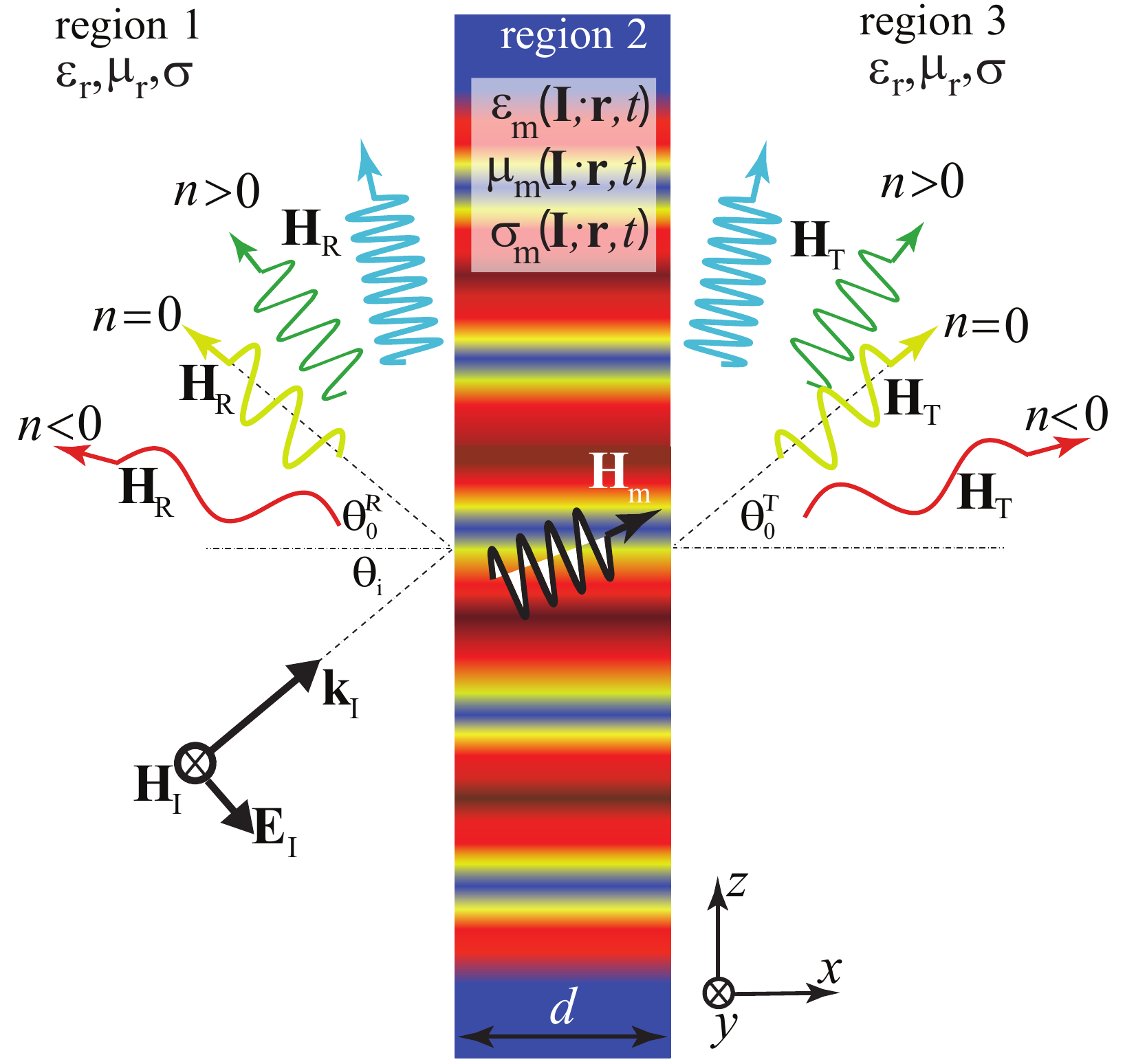}
		\caption{Incidence and scattering of a TM electromagnetic wave from a general space-time-varying medium.}
		\label{Fig:sch_TM}
	\end{center}
\end{figure}
Consider the oblique incidence of a TM electromagnetic wave, as shown in Fig.~\ref{Fig:sch_TM}, where 
\begin{subequations}
	\begin{equation}\label{eqa:F1}
\mathbf{H}= \mathbf{\hat{y}} \mathbf{H}_y(x,z,t),
	\end{equation}
 and 
 	\begin{equation}\label{eqa:F2}
 \mathbf{E}=-\eta \mathbf{\hat{k}} \times \mathbf{H} =	\hat{x} \mathbf{E}_x (x,z,t)  - \mathbf{\hat{z}}  \mathbf{E}_z (x,z,t).
 	\end{equation}
 \end{subequations}
 
  Therefore, Eqs.~\eqref{eqa:Max1} and~\eqref{eqa:Max2} yield
\begin{subequations}
\begin{equation}\label{eqa:Maxw0}
	\mu(\textbf{I};\textbf{r},t)\frac{\partial H_y}{\partial t}+H_y \frac{\partial \mu(\textbf{I};\textbf{r},t) }{\partial t}=  \frac{\partial E_z}{\partial x} - \frac{\partial E_x}{\partial z} ,
\end{equation}
\begin{equation}\label{eqa:Maxw2}
	\epsilon(\textbf{I};\textbf{r},t)\frac{\partial E_x}{\partial t}+ E_x \frac{\partial \epsilon(\textbf{I};\textbf{r},t)}{\partial t}= -   \frac{\partial H_y}{\partial z}-\sigma(\textbf{I};\textbf{r},t) E_x  , 
\end{equation}
\begin{equation}\label{eqa:Maxw3}
\epsilon(\textbf{I};\textbf{r},t)	\frac{\partial E_z}{\partial t} + E_z \frac{\partial \epsilon(\textbf{I};\textbf{r},t)}{\partial t}=   \frac{\partial H_y}{\partial x}-\sigma(\textbf{I};\textbf{r},t) E_z  . 
\end{equation}
\end{subequations}

The finite-difference discretized form of the first two Maxwell's equations for the electric and magnetic fields will be simplified to
\begin{subequations}
\begin{equation}\label{eqa:num_Max1}
		\begin{split}
			E_x\Big\lvert\substack{{k+1/2}\\{j+1/2}\\{i+1/2}}=&\left(1-\Delta t \dfrac{\epsilon' +\sigma }{\epsilon} \Big\lvert\substack{{k}\\{j+1/2}\\{i+1/2}}  \right)	E_x\Big\lvert\substack{{k-1/2}\\{j+1/2}\\{i+1/2}}- \dfrac{\Delta t}{ \epsilon \Big\lvert\substack{{k}\\{j+1/2}\\{i+1/2}}}  \dfrac{H_y\Big\lvert\substack{{k}\\{j+1}\\{i+1}}-H_y\Big\lvert\substack{{k}\\{j}\\{i+1}} }{\Delta z}, 
					\end{split}	
\end{equation}
	\begin{equation}\label{eqa:num_Max2}
		\begin{split}
			E_z\Big\lvert\substack{{k+1/2}\\{j+1/2}\\{i+1/2}}=&\left(1-\Delta t \dfrac{\epsilon'+\sigma }{\epsilon } \Big\lvert\substack{{k}\\{j+1/2}\\{i+1/2}}  \right)	E_z\Big\lvert\substack{{k-1/2}\\{j+1/2}\\{i+1/2}}+ \dfrac{\Delta t}{ \epsilon  \Big\lvert\substack{{k}\\{j+1/2}\\{i+1/2}}  } \dfrac{H_y\Big\lvert\substack{{k}\\{j+1}\\{i+1}}-H_y\Big\lvert\substack{{k}\\{j+1}\\{i}}}{\Delta x}  , 
		\end{split}	
	\end{equation}
	\begin{equation}\label{eqa:num_Max3}
		\begin{split}
			H_y &\Big\lvert\substack{{k+1}\\{j}\\{i}}=\left(1-\Delta t \dfrac{\mu'}{\mu}\Big\lvert\substack{{k+1/2}\\{j}\\{i}}  \right) H_y \Big\lvert\substack{{k}\\{j}\\{i}}- \dfrac{\Delta t}{ \mu \Big\lvert\substack{{k+1/2}\\{j}\\{i}}}  \left( \dfrac{E_x \Big\lvert\substack{{k+1/2}\\{j+1/2}\\{i+1/2}}-E_x\Big\lvert\substack{{k+1/2}\\{j-1/2}\\{i+1/2}}}{\Delta z}  
			- \dfrac{E_z \Big\lvert\substack{{k+1/2}\\{j+1/2}\\{i+1/2}}-E_z\Big\lvert\substack{{k+1/2}\\{j+1/2}\\{i-1/2}} }{\Delta x} \right). 
		\end{split}	
	\end{equation}
where $\epsilon'$ and $\mu'$ are time derivatives of the permittivity and permeability, respectively, and are computed as
	\begin{equation}\label{eqa:der}
\epsilon'\Big\lvert\substack{{k}\\{j+1/2}\\{i+1/2}}=\dfrac{\epsilon\Big\lvert\substack{{k}\\{j+1/2}\\{i+1/2}}-\epsilon\Big\lvert\substack{{k-1}\\{j+1/2}\\{i+1/2}}}{\Delta t}, \qquad  \mu'\Big\lvert\substack{{k+1/2}\\{j}\\{i}}=\dfrac{\mu\Big\lvert\substack{{k+1/2}\\{j}\\{i}}-\mu\Big\lvert\substack{{k-1}\\{j}\\{i}}}{\Delta t}.
	\end{equation}
\end{subequations}
\begin{figure*}
	\begin{center}
		\includegraphics[width=0.8\columnwidth]{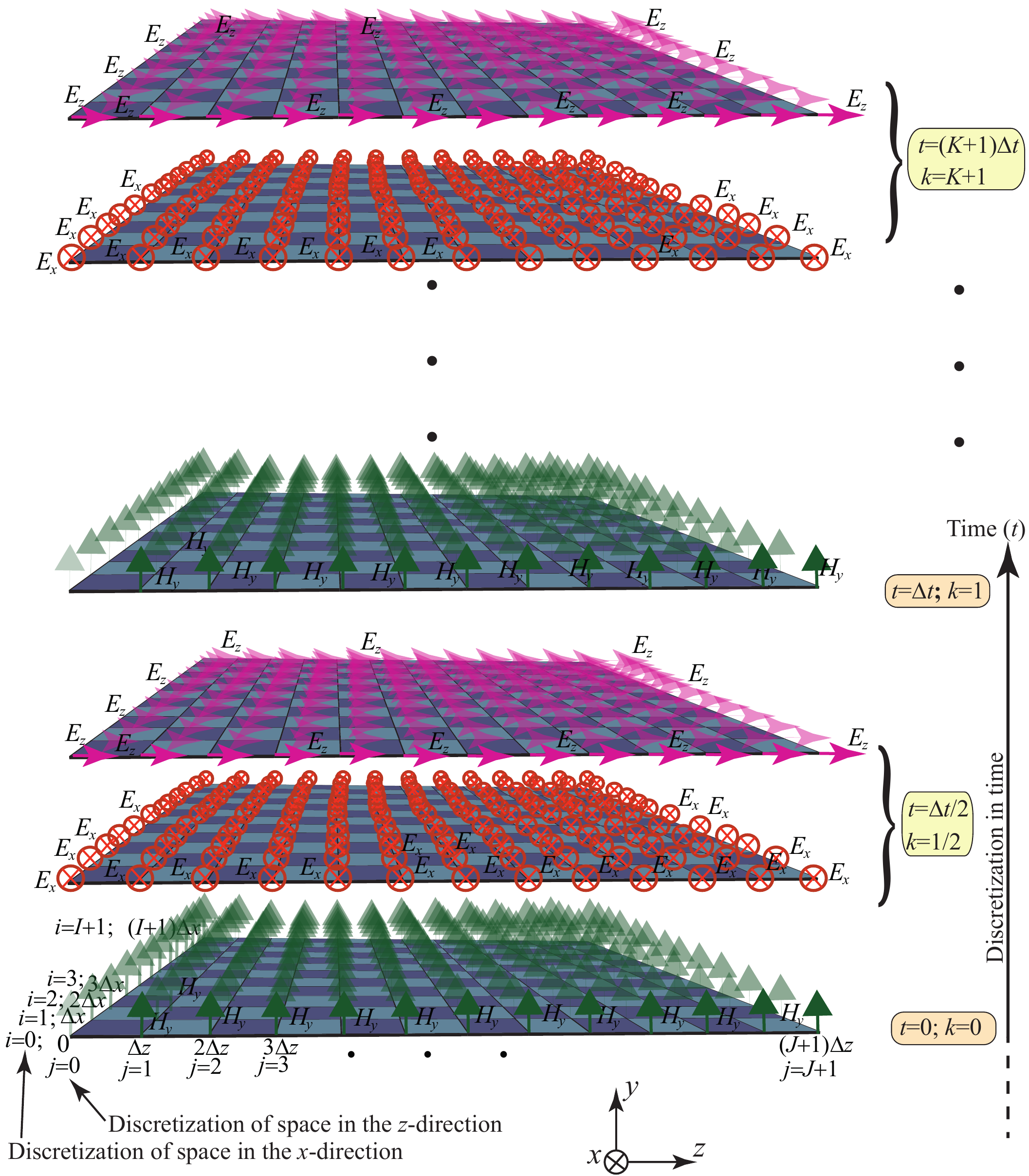} 
		\caption{Finite-difference time-domain scheme for TM wave illumination of a space-time-varying slab.} 
		\label{Fig:FDTD_TM}
	\end{center}
\end{figure*}

Figure~\ref{Fig:FDTD_TM} plots the implemented finite-difference time-domain scheme for numerical simulation of the oblique wave impinging on a space-time-varying medium, where the medium's electrical permittivity, magnetic permeability and electrical conductivity are all modulated in both space and time. We first discretize the medium to $J+1$ spatial samples and $K+1$ temporal samples, with the steps of $\Delta z$ and $\Delta t$, respectively. The angle of incidence is specified by the incident wave source as
\begin{equation} \label{eq:Hy_source}
	H_y\Big\lvert\substack{{k\Delta_t}\\{z_\text{s}}\\{i_1:i_2}} = \cos \left(\omega_0 \cdot k\Delta t  -   \beta_0 (i_1:i_2)\Delta z  \sin( \theta ) \right).
\end{equation}
where $i_1$, $i_2$ and $z_\text{s}$ specify the source location.

To mitigate boundary reflections in the FDTD simulations of space-time-varying metamaterials, Mur's first-order Absorbing Boundary Condition (ABC) is applied. This technique allows outgoing waves to exit the computational domain with minimal reflections, creating a quasi-infinite environment. Mur's boundary conditions involve specific coefficients that dictate how the boundary fields are updated over time. The Mur's absorbing boundary condition parameters are
\begin{subequations}\label{eq:BCs}
\begin{equation}
	c_0 = \frac{c}{2S} \left(1 + \frac{p_0}{S}\right); \quad c_1 = -\frac{c}{2S} \left(1 - \frac{p_0}{S}\right); c_2 = -\frac{c}{S^2} \left(p_0 + p_2 S^2\right); \quad c_3 = \frac{p_2 c}{2}
\end{equation}
\begin{equation}
c_0^{\text{fw}} = -\frac{c_0}{c_1}, \quad c_2^{\text{fw}} = -\frac{c_2}{c_1}, \quad c_3^{\text{fw}} = -\frac{c_3}{c_1}, \quad c_1^{\text{bw}} = -\frac{c_1}{c_0}, \quad c_2^{\text{bw}} = -\frac{c_2}{c_0}, \quad c_3^{\text{bw}} = -\frac{c_3}{c_0}
\end{equation}

where $S=1/\sqrt{2}$, $p_0 = 1$ and $p_2 = -0.5$. At each time step \( k \), the magnetic field component \( H_y \) is updated at the forward, backward, upward and downward boundary cells as
\begin{equation}
	H_y\Big\lvert\substack{{k}\\{I-2}\\{j}}=c_0^{\text{f}} \left( H_y\Big\lvert\substack{{k}\\{I-3}\\{j}}
	+H_y\Big\lvert\substack{{k-2}\\{I-2}\\{j}} \right)- H_y\Big\lvert\substack{{k-2}\\{I-3}\\{j}}
	+ c_2^{\text{f}} \left(H_y\Big\lvert\substack{{k-1}\\{I-2}\\{j}}+H_y\Big\lvert\substack{{k-1}\\{I-3}\\{j}} \right) 
	+c_3^{\text{f}} \left(H_y\Big\lvert\substack{{k-1}\\{I-2}\\j_\text{m}} +H_y\Big\lvert\substack{{k-1}\\{I-3}\\j_\text{p}}+H_y\Big\lvert\substack{{k-1}\\{I-2}\\j_\text{p}}+H_y\Big\lvert\substack{{k-1}\\{I-3}\\{j_\text{m}}} \right),
\end{equation}
\begin{equation}
	H_y\Big\lvert\substack{{k}\\{2}\\{j}}=c_1^{\text{bw}} \left( H_y\Big\lvert\substack{{k}\\{3}\\{j}}
	+H_y\Big\lvert\substack{{k-2}\\{2}\\{j}} \right)- H_y\Big\lvert\substack{{k-2}\\{3}\\{j}}
	+ c_2^{\text{bw}} \left(H_y\Big\lvert\substack{{k-1}\\{2}\\{j}}+H_y\Big\lvert\substack{{k-1}\\{3}\\{j}} \right) 
	+c_3^{\text{bw}} \left(H_y\Big\lvert\substack{{k-1}\\{2}\\j_\text{m}} +H_y\Big\lvert\substack{{k-1}\\{3}\\j_\text{p}}+H_y\Big\lvert\substack{{k-1}\\{2}\\j_\text{p}}+H_y\Big\lvert\substack{{k-1}\\{3}\\{j_\text{m}}} \right),
\end{equation}
\begin{equation}
	H_y\Big\lvert\substack{{k}\\{i}\\{J-2}}=c_0^{\text{fw}} \left( H_y\Big\lvert\substack{{k}\\{i}\\{J-3}}
+H_y\Big\lvert\substack{{k-2}\\{i}\\{J-2}} \right)- H_y\Big\lvert\substack{{k-2}\\{i}\\{J-3}}
+ c_2^{\text{fw}} \left(H_y\Big\lvert\substack{{k-1}\\{i}\\{J-2}}+H_y\Big\lvert\substack{{k-1}\\{i}\\{J-3}} \right) 
+c_3^{\text{fw}} \left(H_y\Big\lvert\substack{{k-1}\\i_\text{m}\\{J-2}} +H_y\Big\lvert\substack{{k-1}\\i_\text{p}\\{J-3}}+H_y\Big\lvert\substack{{k-1}\\i_\text{p}\\{J-2}} +H_y\Big\lvert\substack{{k-1}\\i_\text{m}\\{J-3}} \right),
\end{equation}
\begin{equation}
	H_y\Big\lvert\substack{{k}\\{i}\\{2}}=c_1^{\text{bw}} \left( H_y\Big\lvert\substack{{k}\\{i}\\{3}}
	+H_y\Big\lvert\substack{{k-2}\\{i}\\{2}} \right)- H_y\Big\lvert\substack{{k-2}\\{i}\\{3}}
	+ c_2^{\text{bw}} \left(H_y\Big\lvert\substack{{k-1}\\{i}\\{3}}+H_y\Big\lvert\substack{{k-1}\\{i}\\{2}} \right) 
	+c_3^{\text{bw}} \left(H_y\Big\lvert\substack{{k-1}\\i_\text{m}\\{2}} +H_y\Big\lvert\substack{{k-1}\\i_\text{p}\\{3}}+H_y\Big\lvert\substack{{k-1}\\i_\text{p}\\{2}} +H_y\Big\lvert\substack{{k-1}\\i_\text{m}\\{3}} \right),
\end{equation}
where $i=3:I-3$, $j=3:J-3$, $i_\text{m}=2:I-4$, $i_\text{p}=4:I-2$, $j_\text{m}=2:J-4$ and $j_\text{p}=4:J-2$. For corners, we apply a mirroring technique to account for field propagation, as
\begin{equation}
	H_y\Big\lvert\substack{{k}\\{2}\\{2}}=  H_y\Big\lvert\substack{{k-2}\\{3}\\{3}}; \quad	H_y\Big\lvert\substack{{k}\\{2}\\{J-2}}=  H_y\Big\lvert\substack{{k-2}\\{3}\\{J-3}}; \quad	H_y\Big\lvert\substack{{k}\\{I-2}\\{2}}=  H_y\Big\lvert\substack{{k-2}\\{I-3}\\{3}}; \quad	H_y\Big\lvert\substack{{k}\\{I-2}\\{J-2}}=  H_y\Big\lvert\substack{{k-2}\\{I-3}\\{J-3}}.
\end{equation}
\end{subequations}

To analyze the signal's frequency components, we transform the output time-domain signal into its frequency-domain representation using the Fast Fourier Transform (FFT), as
\begin{equation}\label{eqa:FFT}
	E(\omega)=20 \log_{10} \left| \mathcal{F}_{\text{shift}} \left( \mathcal{F}[E_{\text{out}}(t)] \right) \right| 
\end{equation}
which provides the spectral content of the output signal $E_{\text{out}} $. The process begins with computing the Fast Fourier Transform (FFT) of $E_{\text{out}} $, transforming the signal from the time domain to the frequency domain. Following this, we apply a frequency shift using the FFT shift, centering the zero-frequency component in the spectrum, which facilitates better visualization and analysis. The magnitude of the resulting complex frequency components is then calculated, converting them into real-valued amplitudes. To represent the data in a more perceptually meaningful way, the scaled magnitudes are converted into a logarithmic scale using the base-10 logarithm. Finally, the factor of 20 is applied to express these values in decibels (dB), providing a logarithmic measure of the signal's frequency content. This approach allows for a comprehensive and intuitive understanding of the signal's behavior in the frequency domain, highlighting the relative strengths of various frequency components in a manner that aligns with human auditory perception.

\section{FDTD Scheme and Equations for TE Wave Excitation}\label{sec:TE}
\begin{figure}
	\begin{center}
		\includegraphics[width=0.4\columnwidth]{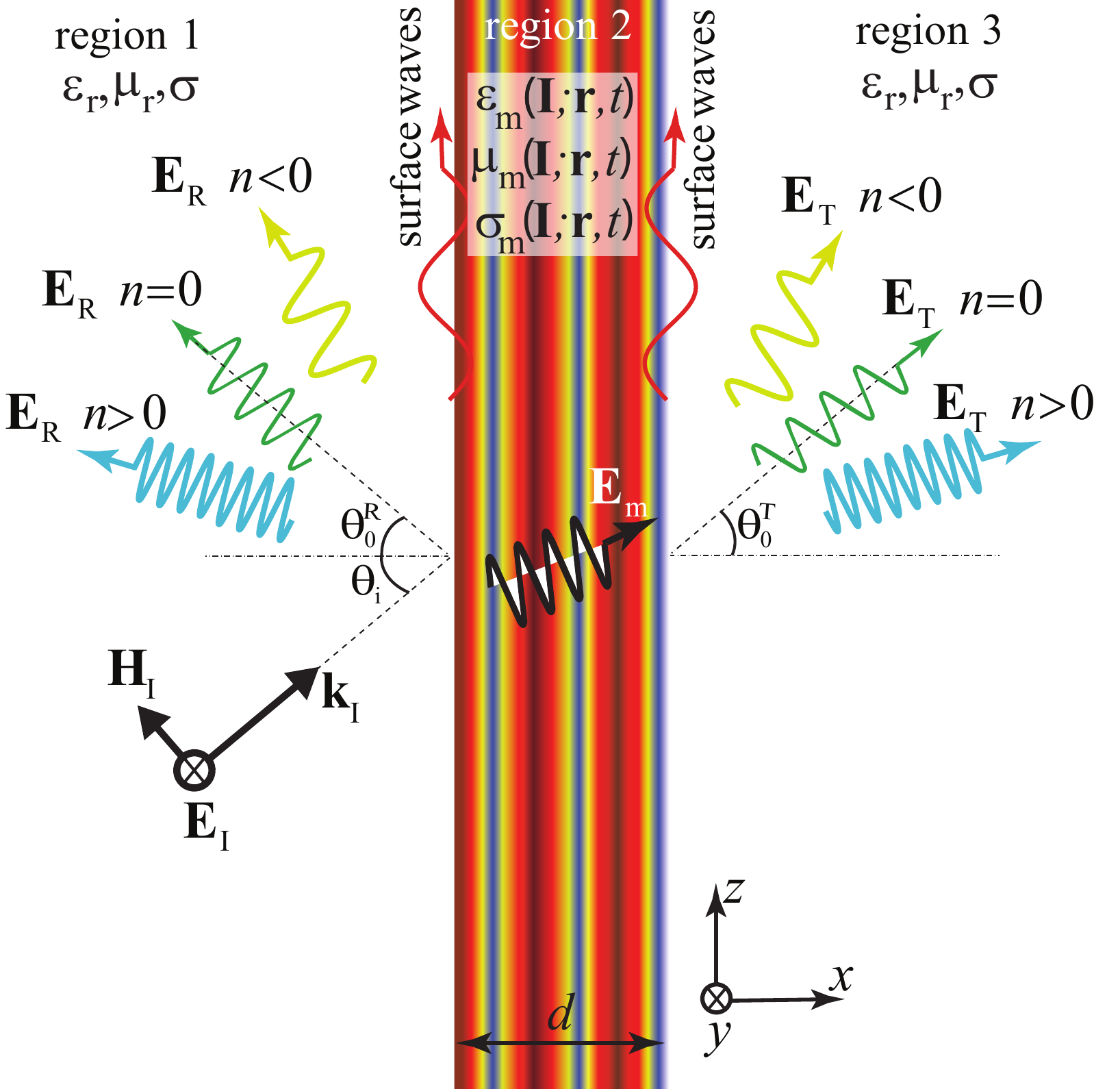}
		\caption{TE wave scattering from a space and time-varying slab.}
		\label{Fig:sch2}
	\end{center}
\end{figure}

Next, we consider the oblique incidence of a TE wave which is depicted in Fig.~\ref{Fig:sch2}. The electric and magnetic fields read 
\begin{subequations}
	\begin{equation}\label{eqa:F3}
		\mathbf{E}= \mathbf{\hat{y}} \mathbf{E}_y(x,z,t),
	\end{equation}
	and 
	\begin{equation}\label{eqa:F4}
		\mathbf{H}=\dfrac{1}{\eta} \left[\mathbf{\hat{k}} \times \mathbf{E}\right]=-	\hat{x} \mathbf{H}_x (x,z,t) + \mathbf{\hat{z}}  \mathbf{H}_z (x,z,t).
	\end{equation}
\end{subequations}

Consequently, Eqs.~\eqref{eqa:Max1} and~\eqref{eqa:Max2} yield
\begin{subequations}
\begin{equation}\label{eqa:Maxw1}
	\epsilon(\textbf{I};\textbf{r},t) \frac{\partial E_y}{\partial t}+ E_y \frac{\partial \epsilon(\textbf{I};\textbf{r},t) }{\partial t}=  \frac{\partial H_x}{\partial z} - \frac{\partial H_z}{\partial x}-\sigma(\textbf{I};\textbf{r},t) E_y ,
\end{equation}
\begin{equation}\label{eqa:Maxw2}
\mu(\textbf{I};\textbf{r},t)\frac{\partial H_x}{\partial t}+ H_x \frac{\partial \mu(\textbf{I};\textbf{r},t)}{\partial t}=   \frac{\partial E_y}{\partial z}, 
\end{equation}
\begin{equation}\label{eqa:Maxw3}
\mu(\textbf{I};\textbf{r},t)	\frac{\partial H_z}{\partial t}+H_z \frac{\partial \mu(\textbf{I};\textbf{r},t)}{\partial t}= -\frac{\partial E_y}{\partial x}. 
\end{equation}
	\end{subequations}
		
\begin{figure*}
	\begin{center}
		\includegraphics[width=0.8\columnwidth]{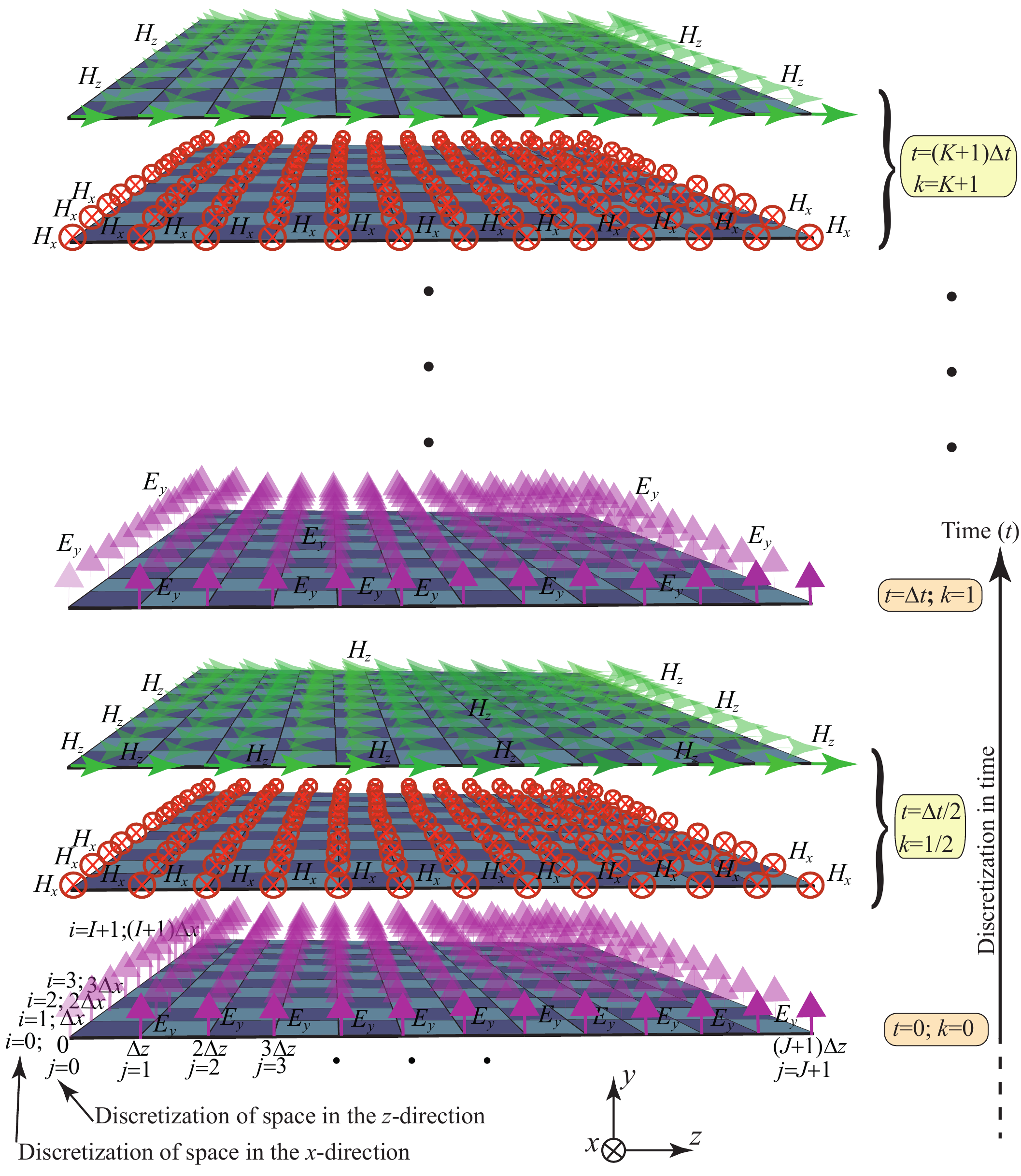} 
		\caption{Finite-difference time-domain scheme for oblique incidence of TE wave to a space-time-modulated slab.} 
		\label{Fig:FDTD_TE}
	\end{center}
\end{figure*}

 Figure~\ref{Fig:FDTD_TE} plots the implemented finite-difference time-domain scheme for numerical simulation of the oblique TE wave impinging on a space-time-varying medium, where the electrical permittivity, magnetic permeability and electrical conductivity of the medium are all modulated in both space and time. The finite-difference discretized form of the first two Maxwell's equations for the electric and magnetic fields read
\begin{subequations}
	\begin{equation}\label{eqa:num_Max1c}
	\begin{split}
		H_x\Big\lvert\substack{{k+1/2}\\{j+1/2}\\{i+1/2}}=&	\left(1-\Delta t \dfrac{\mu'}{\mu}\Big\lvert\substack{{k}\\{j+1/2}\\{i+1/2}}   \right)   H_x\Big\lvert\substack{{k-1/2}\\{j+1/2}\\{i+1/2}}+\dfrac{\Delta t}{ \mu \Big\lvert\substack{{k}\\{j+1/2}\\{i+1/2}} }  \dfrac{E_y\Big\lvert\substack{{k}\\{j+1}\\{i}}-E_y\Big\lvert\substack{{k}\\{j}\\{i}}}{\Delta z}  , 
	\end{split}	
\end{equation}
\begin{equation}\label{eqa:num_Max1d}
	\begin{split}
		H_z\Big\lvert\substack{{k+1/2}\\{j+1/2}\\{i+1/2}}=&	\left(1-\Delta t \dfrac{\mu'}{\mu}\Big\lvert\substack{{k}\\{j+1/2}\\{i+1/2}}    \right)  H_z\Big\lvert\substack{{k-1/2}\\{j+1/2}\\{i+1/2}}-\dfrac{\Delta t}{ \mu \Big\lvert\substack{{k}\\{j+1/2}\\{i+1/2}}  }  \dfrac{E_y\Big\lvert\substack{{k}\\{j}\\{i+1}}-E_y\Big\lvert\substack{{k}\\{j}\\{i}}}{\Delta x}  , 
	\end{split}	
\end{equation}
\begin{equation}\label{eqa:num_Max2c}
		\begin{split}
			E_y& \Big\lvert\substack{{k+1}\\{j}\\{i}}= \left(1-\Delta t  \dfrac{\epsilon'+ \sigma }{\epsilon} \Big\lvert\substack{{k+1/2}\\{j}\\{i}}  \right) E_y \Big\lvert\substack{{k}\\{j}\\{i}}+ \dfrac{\Delta t}{ \epsilon\Big\lvert\substack{{k+1/2}\\{j}\\{i}} } 
			\left(\dfrac{H_x \Big\lvert\substack{{k+1/2}\\{j+1/2}\\{i+1/2}}-H_x\Big\lvert\substack{{k+1/2}\\{j-1/2}\\{i+1/2}}}{\Delta z}  -   \dfrac{H_z \Big\lvert\substack{{k+1/2}\\{j+1/2}\\{i+1/2}}-H_z \Big\lvert\substack{{k+1/2}\\{j+1/2}\\{i-1/2}}}{\Delta x}\right). 
		\end{split}	
	\end{equation}
where
\begin{equation}\label{eqa:der2}
	\mu'\Big\lvert\substack{{k}\\{j+1/2}\\{i+1/2}}=\dfrac{\mu\Big\lvert\substack{{k}\\{j+1/2}\\{i+1/2}}-\mu\Big\lvert\substack{{k-1}\\{j+1/2}\\{i+1/2}}}{\Delta t}, \qquad  \epsilon'\Big\lvert\substack{{k+1/2}\\{j}\\{i}}=\dfrac{\epsilon\Big\lvert\substack{{k+1/2}\\{j}\\{i}}-\epsilon\Big\lvert\substack{{k-1}\\{j}\\{i}}}{\Delta t}.
\end{equation}
\end{subequations}

We apply Mur's first-order absorbing boundary condition from~\eqref{eq:BCs} applying it to $E_y$ (in place of $H_y$) to mitigate boundary reflections in the FDTD simulations of space-time-varying metamaterials. Furthermore, a source field is applied for excitation of the $E_y$ field, similar to the $H_y$ source field in~\eqref{eq:Hy_source}.

\subsection{Illustrative Examples}

\subsubsection{Multifunctional Operation}
\begin{figure}
	\begin{center}
		\includegraphics[width=0.5\columnwidth]{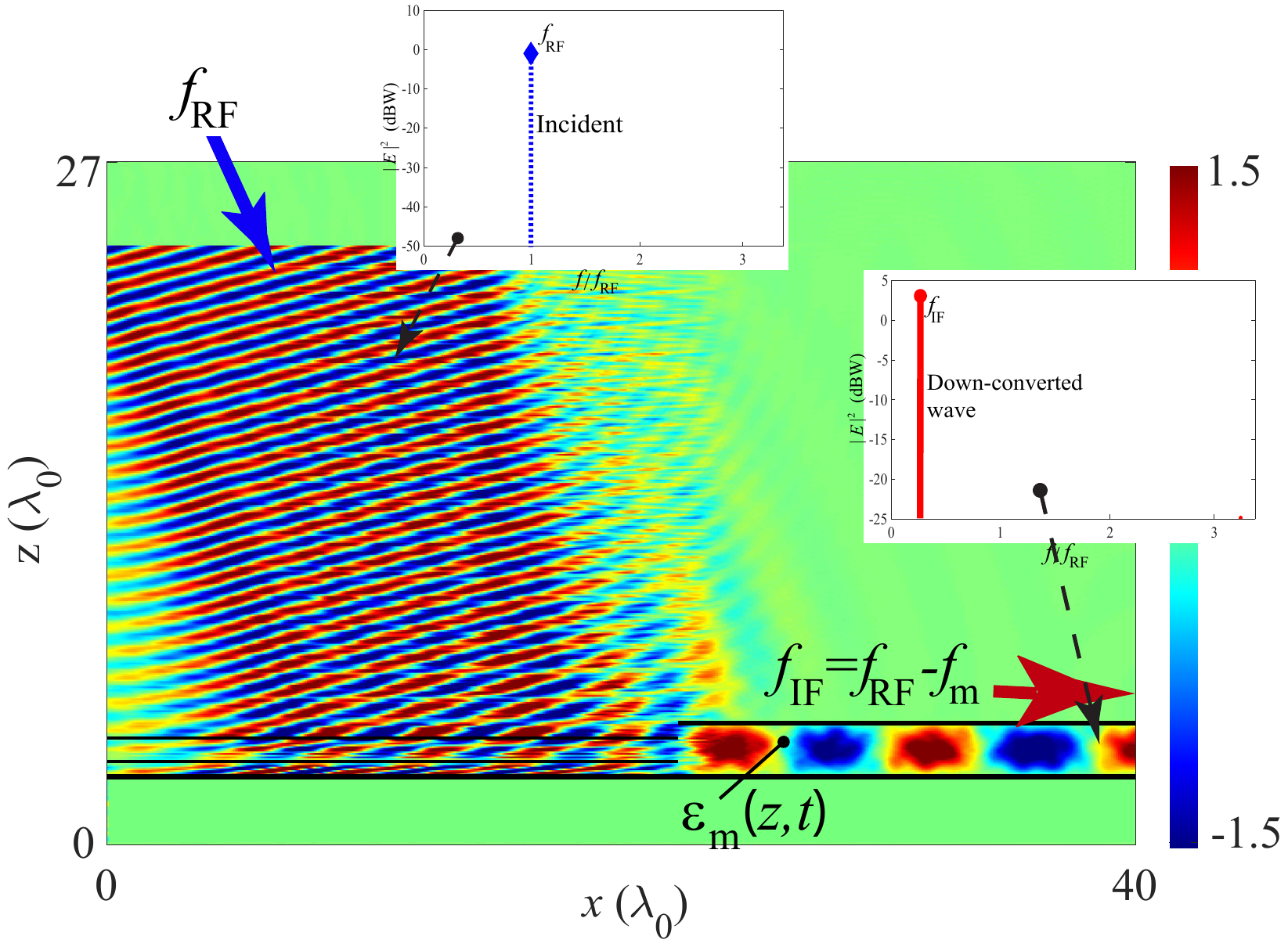} 
		\caption{FDTD time-domain 2D results and frequency spectrum for antenna-mixer-amplifier functionality of space-time-modulated permittivity slab.} 
		\label{Fig:ant}
	\end{center}
\end{figure}

Next, we investigate the effect of TE wave illumination on a space-time-varying permittivity described by
\begin{equation}\label{eqa:ap_permit}
		\epsilon_\text{m}(z,t)= \epsilon_0 \epsilon_\text{r} \left(1+ \delta_\epsilon \cos[\beta_\text{m} z-\omega_\text{m} t+\phi]\right),
\end{equation}
where $\mu$ and $\sigma$ of the slab remain invariant with respect to space and time. In Fig.~\ref{Fig:ant}, we depict the oblique wave incidence and transmission from the slab, with the permittivity described by Equation~\eqref{eqa:ap_permit}, where $\omega_\text{m}>\omega_0$, leading to the intriguing functionality resembling an antenna-mixer-amplifier~\cite{Taravati_AMA_PRApp_2020}. This setup facilitates the transition of space-time harmonics from a propagating wave to a space-time surface wave at a lower frequency within the modulated medium.

\subsubsection{Unidirectional Generation of Space-Time Harmonics}

\begin{figure}
	\begin{center}
		\subfigure[]{\label{Fig:f}
			\includegraphics[width=0.42\linewidth]{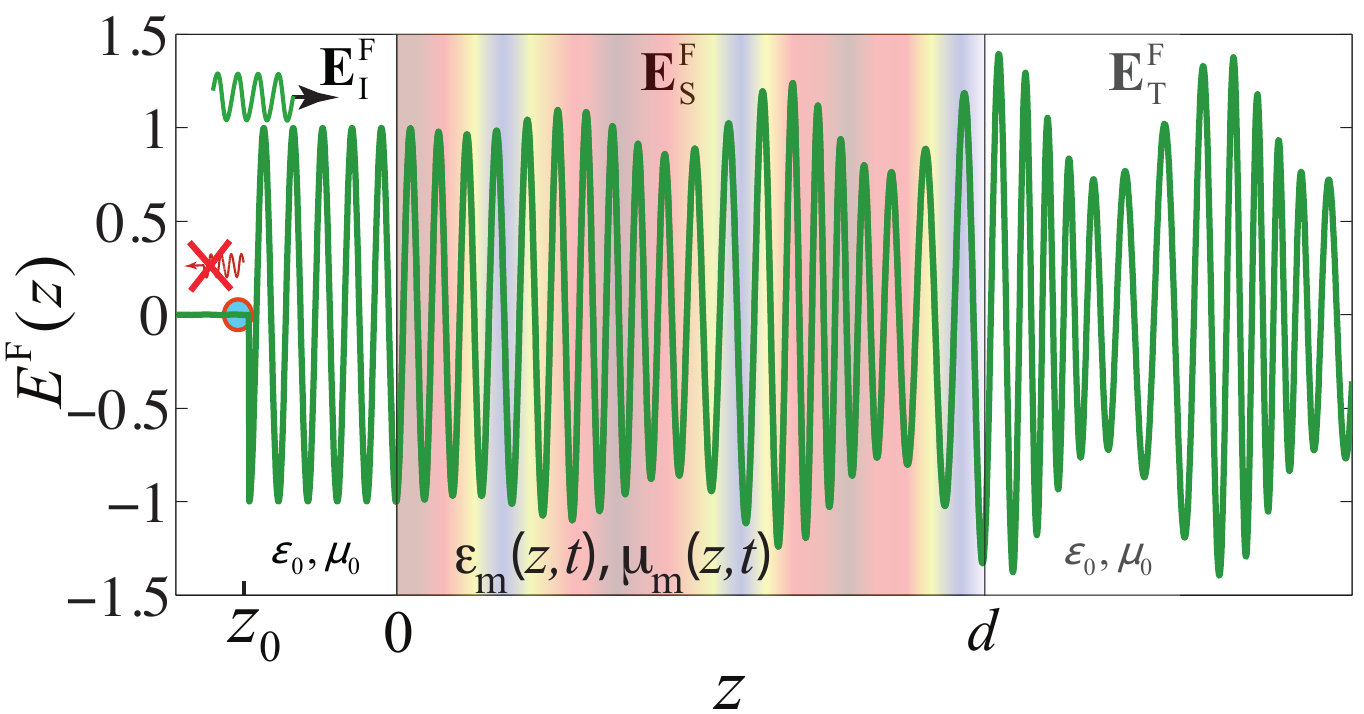}  }
		\subfigure[]{\label{Fig:b}
			\includegraphics[width=0.42\linewidth]{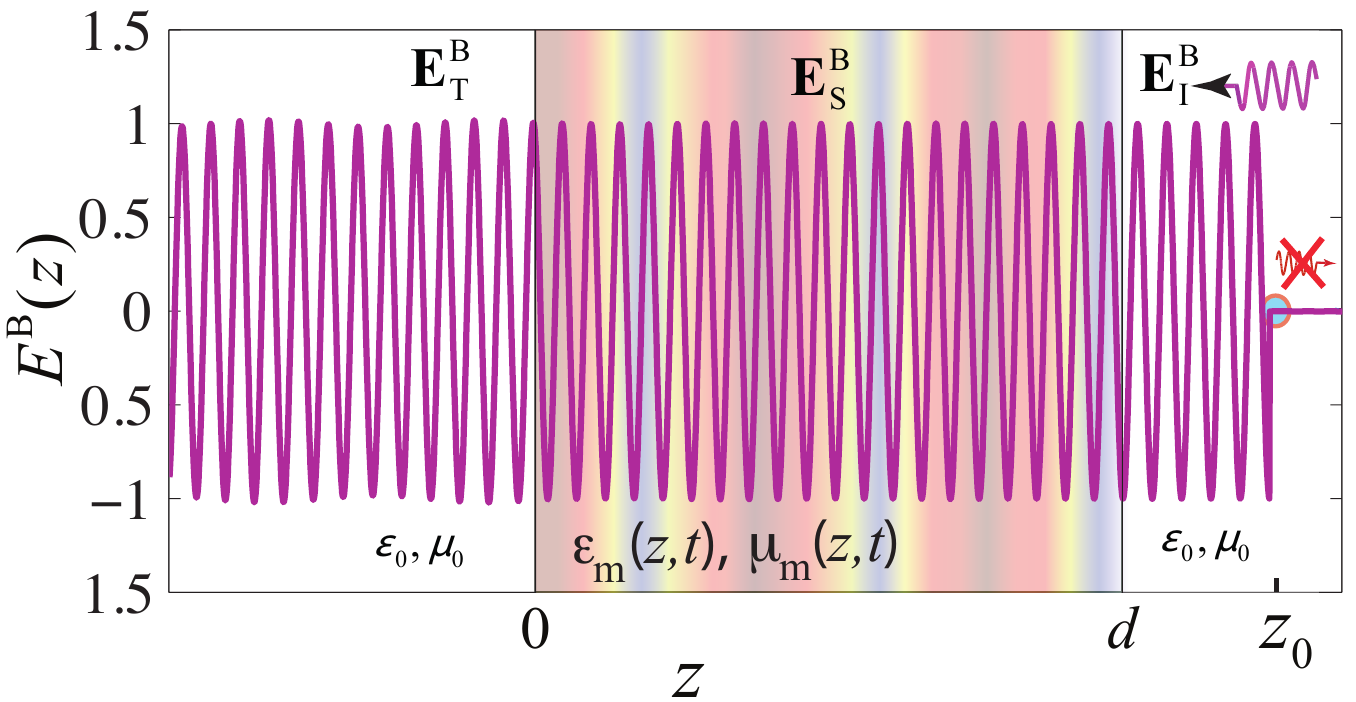}  }
		\caption{FDTD time-domain 1D results for for normal incidence demonstrating nonreciprocal wave transmission from a space-time-modulated slab. (a) Forward wave transmission. (b) Backward wave transmission.}
		\label{Fig:nr}
	\end{center}
\end{figure}

To demonstrate nonreciprocal and asymmetric wave transmission of space-time-varying slabs, we investigate the effect of TE wave illumination on a linear permittivity-and-permeability space-time-varying described space-time-varying permittivity in~\eqref{eqa:ap_permit} and space-time-varying permeability:
\begin{equation}\label{eqa:ap_permeab2}
	\mu_\text{m}(z,t)= \mu_0 \mu_\text{r} \left(1+ \delta_\mu \cos[\beta_\text{m} z-\omega_\text{m} t+\phi]\right),
\end{equation}
with the conductivity ($\sigma$) of the slab remaining invariant with respect to space and time. Figures~\ref{Fig:f} and~\ref{Fig:b} present the result of the 1D-space FDTD investigation into the nonreciprocal wave transmission characteristics from the slab. Here, $\omega_\text{m}<<\omega_0$, leading to the generation of space-time harmonics in the forward case (Fig.\ref{Fig:f}), while no such generation occurs in the backward case (Fig.\ref{Fig:b}).

\begin{figure}
	\begin{center}
	\subfigure[]{\label{Fig:BSa}
			\includegraphics[width=0.48\columnwidth]{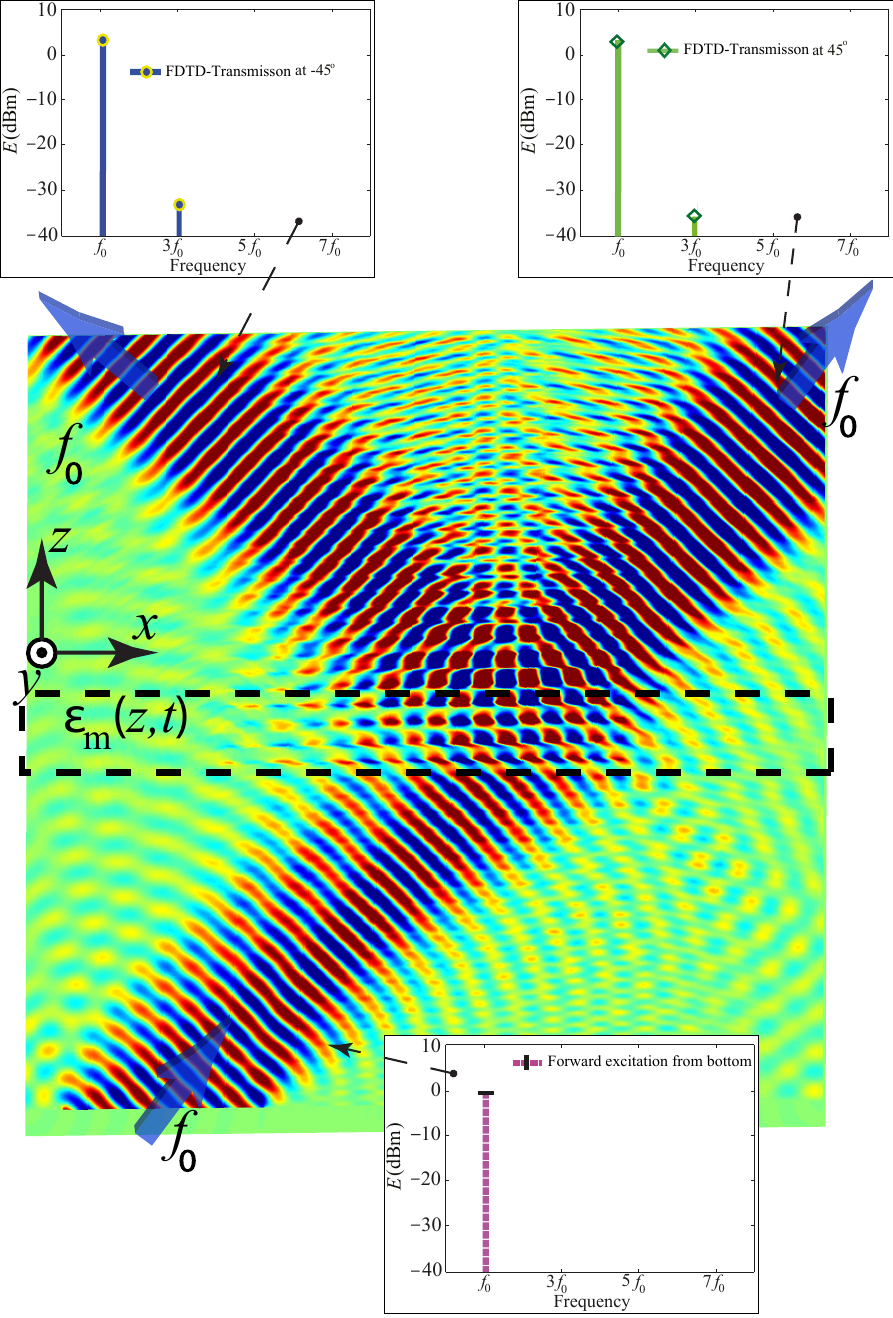} }
				\subfigure[]{\label{Fig:BSb}
				\includegraphics[width=0.48\columnwidth]{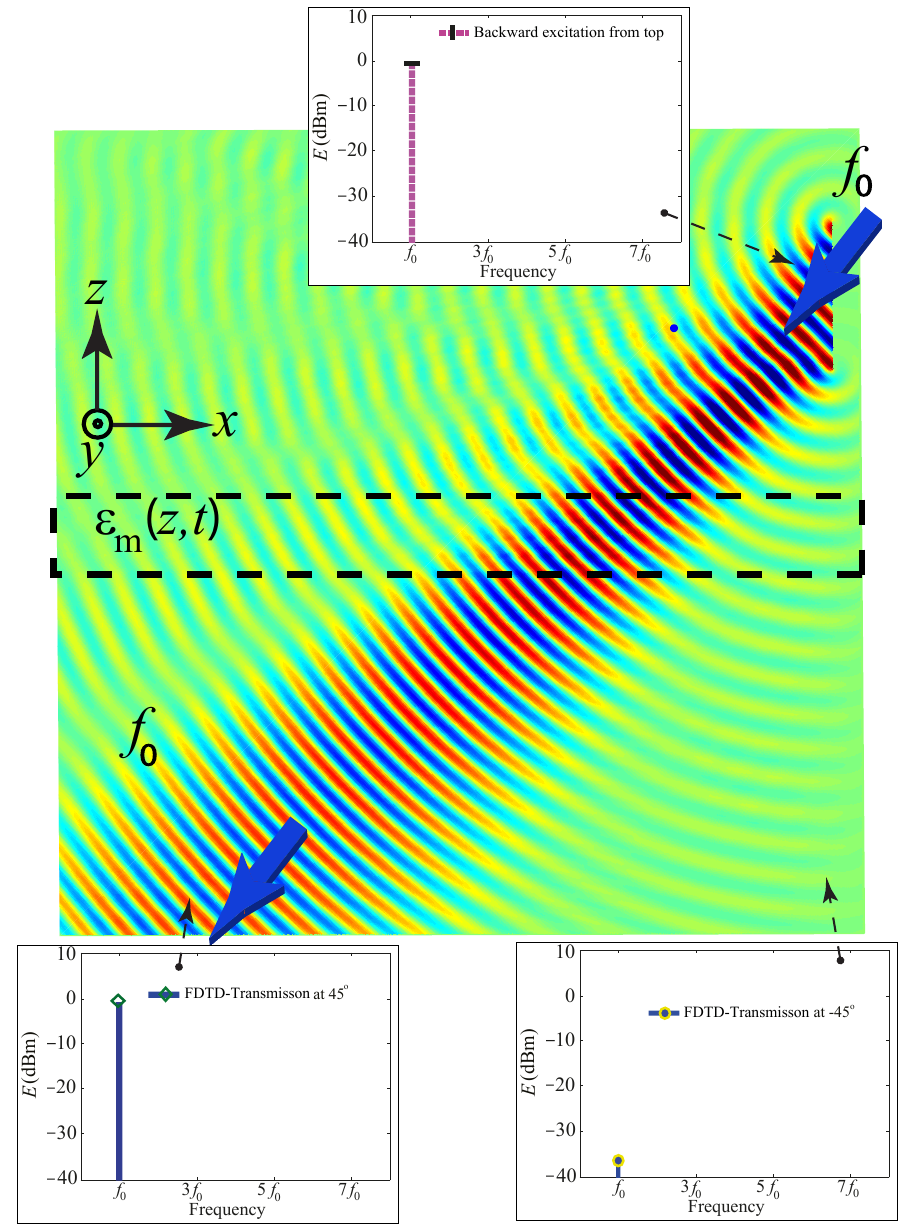} }
		\caption{FDTD time-domain 2D results and frequency spectrum for beam splitting functionality of a space-time-modulated permittivity slab. (a)  Forward excitation from the bottom-left. (b)  Backward excitation from the top-right.} 
		\label{Fig:bs}
	\end{center}
\end{figure}

Figures~\ref{Fig:BSa} and~\ref{Fig:BSb} showcase the unidirectional beam-splitting functionality of space-time-varying slabs with the permittivity defined by Eq.~\eqref{eqa:ap_permit}, where $\omega_\text{m}=2\omega_0$~\cite{Taravati_Kishk_PRB_2018}. Here, the incident wave from the bottom is transmitted while undergoing beam splitting. Consequently, the transmitted beam retains the same frequency as the incident wave, namely  $\omega_0$.

\subsubsection{Space-Time-Modulated Diffraction Grating}
\begin{figure*}
	\begin{center}
		\subfigure[]{\label{Fig:diff1}
			\includegraphics[width=0.98\columnwidth]{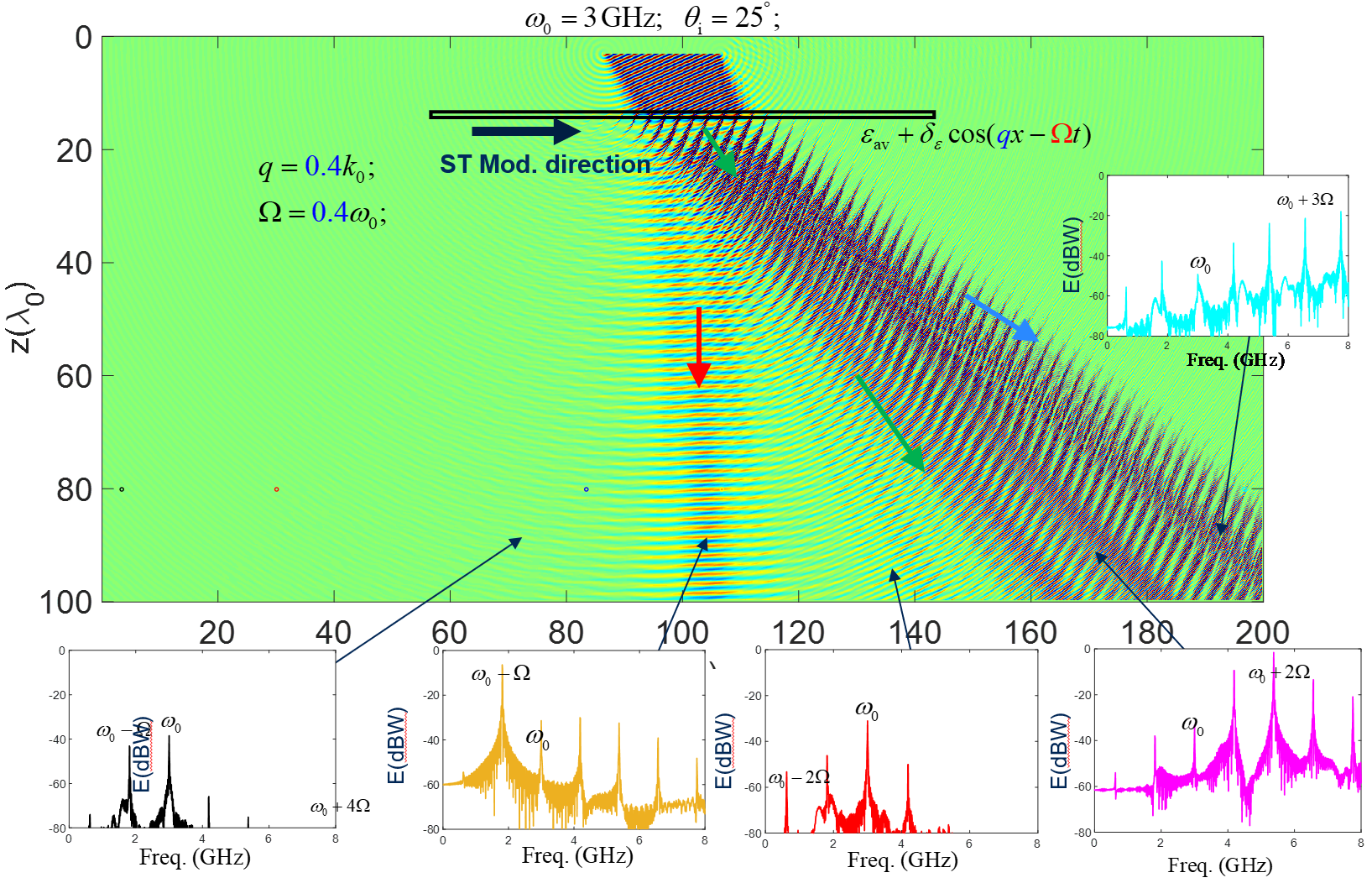} }
		\subfigure[]{\label{Fig:diff2}
			\includegraphics[width=0.98\columnwidth]{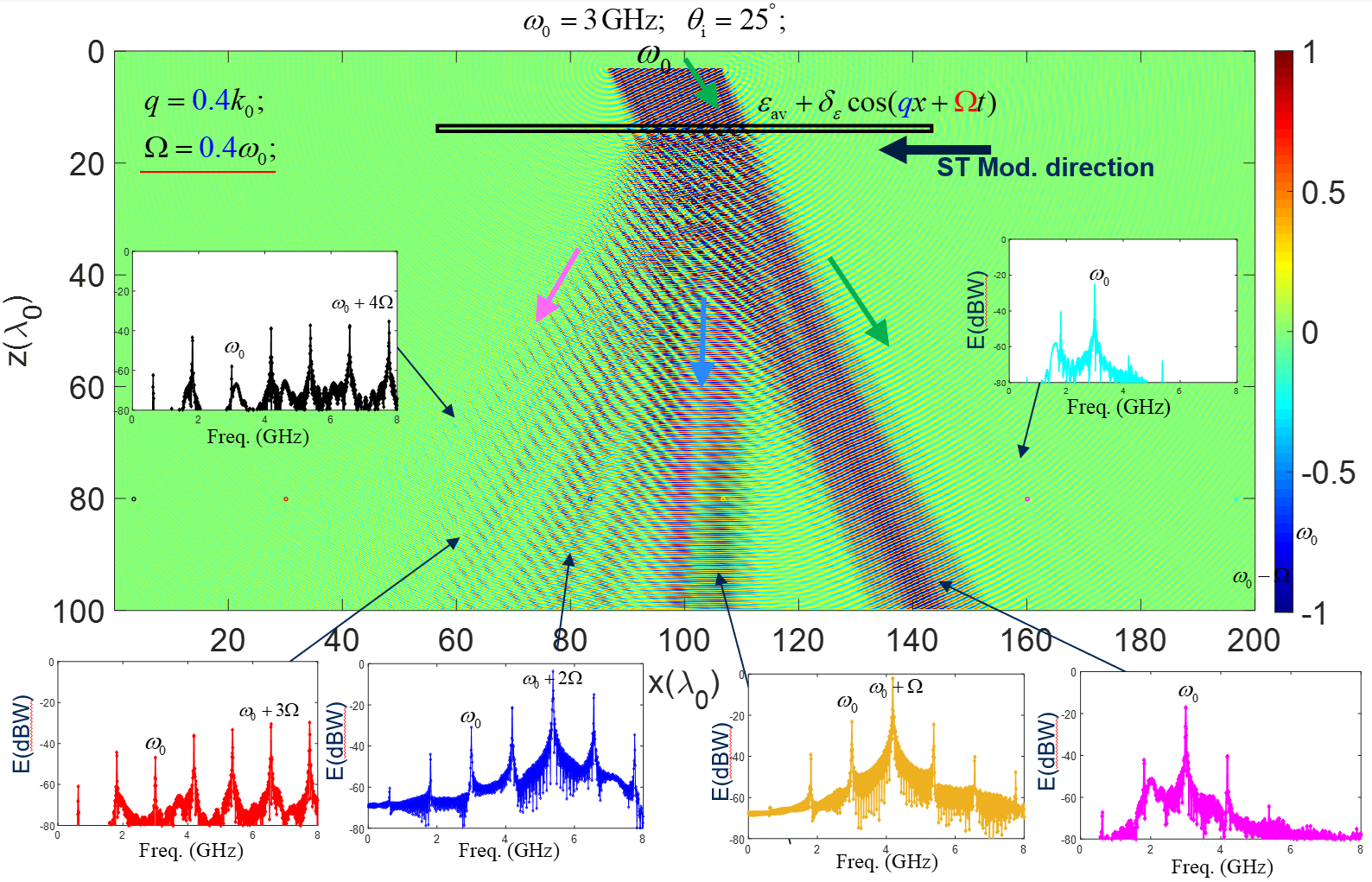} }
		\caption{FDTD time-domain 2D results and frequency spectrum for nonreciprocal spatiotemporal diffraction from a space-time-periodic permittivity-modulated diffraction grating: (a)~Modulation travels along the $+x$ direction, from left to right. (b)~Modulation travels along the $-x$ direction, from right to left.} 
		\label{Fig:diff}
	\end{center}
\end{figure*}

Fig.~\ref{Fig:diff1} shows FDTD time-domain and frequency spectrum results for wave diffraction from a space-time-periodic grating with traveling space-time permittivity modulation~\cite{taravati_PRApp_2019}, where the modulation travels along the \(+x\) direction, from left to right. The 2D time-domain results in Fig.~\ref{Fig:diff1} clearly demonstrate that spatial diffractions are directed to the right side of the grating. This indicates that the traveling modulation in the \(+x\) direction imparts momentum to the diffracted waves, causing them to propagate in the same direction. The frequency spectrum results, shown as insets in Fig.~\ref{Fig:diff1}, reveal that these spatial diffractions include an infinite number of \(\omega_0 \pm n \Omega\) time diffractions (\(-\infty < n < +\infty\)). Among these, one temporal diffraction is particularly strong and dominant. This dominance suggests that the energy of the incident wave is primarily transferred into this specific temporal harmonic, resulting in a pronounced diffraction peak.

Fig.~\ref{Fig:diff2} presents FDTD time-domain and frequency spectrum results for wave diffraction from a space-time-periodic grating with traveling space-time permittivity modulation, where the modulation travels along the $-x$ direction, from left to right. The 2D time-domain results in Fig.~\ref{Fig:diff2} show that spatial diffractions are directed toward the left side of the main incident wave propagating at an angle of $30^\circ$. This behavior indicates that the traveling modulation in the $-x$ direction induces a momentum transfer in the opposite traveling direction of the modulation travel, causing the diffracted waves to propagate leftward. The frequency spectrum results, shown as insets in Fig.~\ref{Fig:diff2}, indicate that these spatial diffractions also include an infinite number of $\omega_0 \pm n \Omega$ time diffractions. Similar to the previous case, one temporal diffraction is notably stronger and dominant. This suggests a consistent mechanism where the traveling modulation selectively enhances a particular temporal harmonic, concentrating the energy into this dominant diffraction mode.

The FDTD simulations provide insights into asymmetric and nonreciprocal wave diffraction behavior from space-time-periodic gratings with traveling permittivity modulations. When the modulation travels in the $+x$ direction, spatial diffractions are directed to the right, aligning with the modulation direction. Conversely, when the modulation travels in the $-x$ direction, the spatial diffractions are directed leftward, opposite to the modulation direction. In both cases, the frequency spectrum analysis reveals the presence of an infinite number of $\omega_0 \pm n \Omega$ time diffractions. However, one temporal diffraction consistently emerges as stronger and dominant. This dominance can be attributed to the resonant interaction between the incident wave and the traveling modulation, which selectively amplifies a specific temporal harmonic. These findings highlight the critical role of modulation direction in controlling the spatial and temporal characteristics of diffracted waves. Such control mechanisms could be exploited in designing advanced wave manipulation devices for communication systems and photonic circuit applications, including beam steerers, frequency converters, and modulators.



\section{Conclusion}\label{sec:conc}

A comprehensive investigation has been provided into the Finite-Difference Time-Domain (FDTD) numerical modeling of electromagnetic wave transmission through linear and nonlinear space-time-varying media. Through the provided FDTD scheme and electromagnetic field update equations for both TM and TE wave illuminations, researchers and engineers can easily simulate and analyze the behavior of space-time-varying slabs. Moreover, the illustrative examples presented in this paper highlight the diverse functionality and applications of these components. We believe that the insights provided in this paper will inspire further research and development in the field of space-time-varying components, paving the way for innovative solutions in next-generation wireless communication systems, biomedicine, radars, and quantum technologies.

\bibliographystyle{IEEEtran}
\bibliography{Taravati_Reference.bib}

\end{document}